\documentclass[preprint]{revtex4-1}
\usepackage{amssymb}
\usepackage{amsmath}
\usepackage{hyperref}
\usepackage{graphicx}
\usepackage{url} 
\usepackage{outline}
\usepackage{enumitem}
\usepackage{mhchem}
\usepackage{gensymb}
\usepackage[export]{adjustbox}
\usepackage{array}
\usepackage{braket}

\usepackage{listings}

\usepackage{pdfpages}

\makeatletter
\AtBeginDocument{\let\LS@rot\@undefined}
\makeatother

\begin{document}

\title{Characterization of Optical and Spin Properties of Single Tin-Vacancy Centers in Diamond Nanopillars}

\author{Alison E. Rugar}
\author{Constantin Dory} 
\author{Shuo Sun} 
\author{Jelena Vu\v{c}kovi\'c}
\affiliation{E. L. Ginzton Laboratory, Stanford University, Stanford, CA 94305, USA}

\begin{abstract}

Color centers in diamond have attracted much interest as candidates for optically active, solid-state quantum bits. Of particular interest are inversion-symmetric color centers based on group-IV impurities in diamond because they emit strongly into their zero-phonon lines and are insensitive to electric field noise to first order. Early studies of the negatively charged tin-vacancy (SnV$^{-}$) center in diamond have found the SnV$^{-}$ to be a promising candidate: it has high quantum efficiency, emits strongly into its zero-phonon lines, and is expected to have a long $T_2$ spin coherence time at 4~K. To develop the SnV$^{-}$ into a spin qubit requires further characterization, especially of the spin and optical properties of individual SnV$^{-}$ in nanofabricated structures. In this work we isolate single SnV$^{-}$ centers in diamond nanopillars and characterize their emission properties and their spin response to a magnetic field. We observe narrow emission linewidths $<250$~MHz, as well as a strong polarization dependence of each transition. We also find the Zeeman splitting under a magnetic field to be in good agreement with theoretical prediction. Our results pave the way toward future employment of single SnV$^{-}$ centers as optically accessible quantum memories.

\end{abstract}

\maketitle

\section{Introduction}

Color centers in diamond have emerged in recent years as candidates for solid-state, optically active quantum bits (qubits) \cite{AharonovichDiamondCCReview2011,Atature2018,Awschalom2018}. 
A strong spin qubit candidate of this type should have strong emission into a narrow-linewidth transition, good immunity to environmental fluctuations, and long spin coherence time. 
The negatively charged silicon-vacancy (SiV$^{-}$) center, in particular, has garnered a lot of attention in recent years for fulfilling two of these three criteria: strong ($\sim70\%$) emission into its four zero-phonon lines (ZPL) and immunity to electric field noise to first order \cite{LukinIndistinguishableSiV}. However, the SiV$^{-}$ suffers from short coherence times at 4~K ($\sim100$~ns \cite{PingaultSiVB}) and low quantum efficiencies ($\sim10\%$ \cite{SzenesSiVQE}). Thus, other color centers with potentially better optical and spin properties must be explored.

\begin{figure}[h]

\includegraphics[width=0.45\textwidth,]{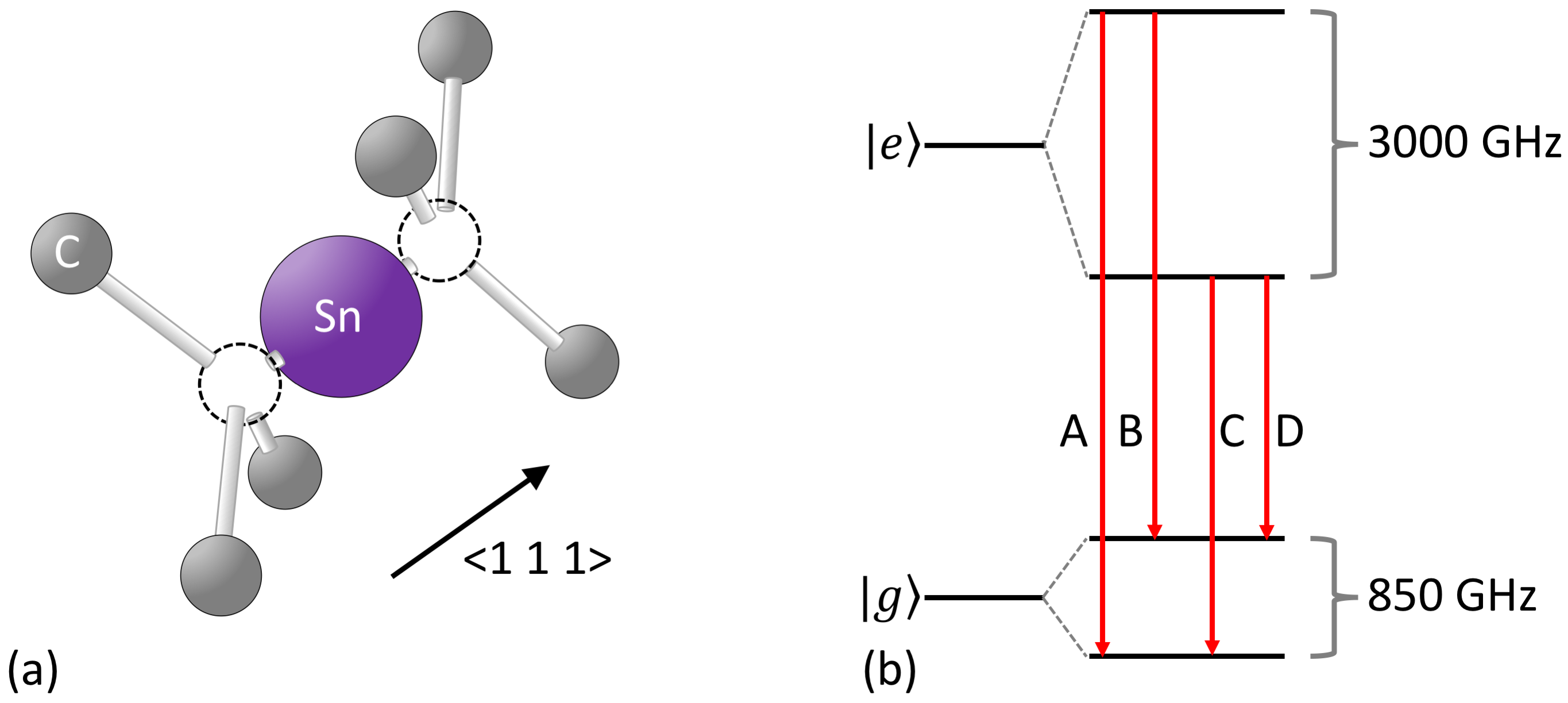} 
\caption{SnV$^{-}$ center in diamond. (a) Schematic of the physical structure of the SnV$^{-}$ center in diamond. Gray spheres represent carbon atoms, the purple sphere represents a tin atom, and the dashed circles outline the vacant carbon lattice sites. The atoms are not exactly to scale. The high-symmetry axis is along $\langle 1 1 1\rangle$. (b) Schematic energy level diagram. The excited- and ground-state manifolds are each split into two states by spin-orbit coupling. In an unstrained SnV$^{-}$, the ground-state splitting is $\sim850$~GHz, and the excited-state splitting is $\sim3000$~GHz. Four optical transitions are allowed, represented by red arrows labeled A, B, C, D in order of decreasing energy.}
\label{SnVstructure_fig}
\end{figure}

The negatively charged tin-vacancy (SnV$^{-}$) center and SiV$^{-}$ belong to the same class of inversion-symmetric color centers comprised of an interstitial group-IV impurity atom accompanied by a split vacancy in the diamond lattice, shown schematically in Fig. \ref{SnVstructure_fig}(a). Thus, SnV$^{-}$ has some of the favorable properties of the SiV$^{-}$ along with some added benefits. 
Notably, preliminary measurements have estimated the quantum efficiency of SnV$^{-}$ to be around $80\%$ \cite{IwasakiSnV}, a significant improvement over that of SiV$^{-}$ $\sim10\%$ \cite{SzenesSiVQE,NeuSiVQE}. Considering that the SnV$^{-}$ has a Debye-Waller factor of $40\%$ \cite{GaliPRX}, the overall probability that an excited SnV$^{-}$ will emit into one of its ZPL is approximately $30\%$, compared to $7\%$ for SiV$^{-}$. 
Furthermore, the SnV$^{-}$ is expected to have improved spin coherence times ($T_2$) over the SiV$^{-}$ \cite{GaliPRX} because of its 17-fold greater ground-state splitting of $\sim850$~GHz (Fig. \ref{SnVstructure_fig}(b)). This larger splitting reduces the probability of single-phonon-mediated transitions between ground states, the primary mechanism of spin decoherence in SiV$^{-}$ \cite{PingaultSiVB}. The SnV$^{-}$ is thus expected to have long $T_2$ at 4~K \cite{GaliPRX}, making the operating conditions of SnV$^{-}$ more feasible.

Central to exploring the SnV$^{-}$ as a qubit candidate is the thorough study of the optical and spin properties of individual SnV$^{-}$ centers. In this work, we study single SnV$^{-}$ centers in diamond nanopillars and characterize their optical and spin properties. In particular, we observe sub-GHz linewidths and a clear polarization dependence of the emission at 5~K. 
We also investigate the behavior of a single SnV$^{-}$ in an external magnetic field at 1.7~K and compare our experimental results to theoretical predictions. Our findings enable further investigations into the spin properties of the SnV$^{-}$ and incorporation of the SnV$^{-}$ into more sophisticated nanophotonic structures \cite{DoryOptimizedDiamondPhotonics}, two crucial steps toward implementation of SnV$^{-}$ as a spin qubit in a quantum network.

\section{Isolation of Single Tin-Vacancy Centers}\label{theory_section}
\begin{figure}[h]
\includegraphics[width=0.45\textwidth,]{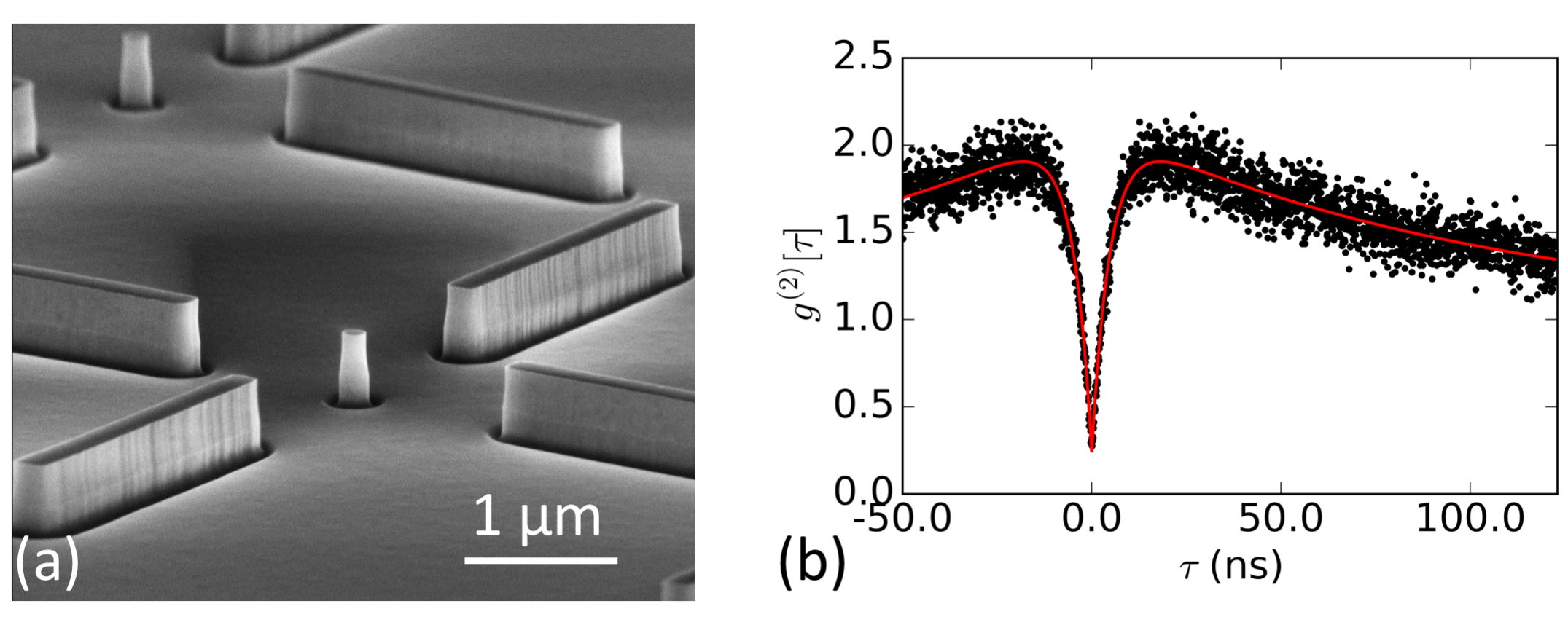} 
\caption{Single SnV$^{-}$ isolated in a nanopillar. (a) SEM image of a 500-nm tall nanopillar etched into implanted and annealed diamond. Neighboring bars are used to help locate pillars. (b) Second-order autocorrelation measurement for a single SnV$^{-}$ in a nanopillar. The data are represented by black dots. The solid red curve is a fit of Eq. \ref{g2_eqn} to the data. From the fit, we find $g^{(2)}[0]=0.23\pm 0.02$.}
\label{pillar_g2_fig}
\end{figure}

Starting with electronic grade, single crystal diamond from Element Six, we perform a boiling tri-acid (1:1:1 nitric:sulfuric:perchloric acids) clean. We then etch 300 nm with oxygen (O$_2$) plasma to alleviate strain and damage near the surface caused by polishing \cite{BurekHighQSiVCavity2014}. This sample is sent to Cutting Edge Ions for $^{120}$Sn$^+$ ion implantation at 370 keV, with a dose of $2\times10^{11}$~cm$^{-2}$. Stopping and Range of Ions in Matter (SRIM) simulations predict a depth of $\sim90$~nm. The implanted diamond is then annealed at 800$\degree$C for 30 minutes and 1100$\degree$C for 90 minutes under vacuum $<10^{-4}$~Torr. 
To make the pillars, we first grow 200~nm of silicon nitride (Si$_x$N$_y$) by low-pressure chemical vapor deposition. Nanopillars, ranging in diameter from 140 nm to 300 nm, are then defined by electron-beam lithography. We transfer the pattern from the developed electron-beam resist into the Si$_x$N$_y$ by a SF$_6$+CH$_4$+N$_2$ plasma etch. Using the patterned Si$_x$N$_y$ as an etch mask, we etch 500~nm into the diamond with a directional O$_2$ plasma etch. A scanning electron microscope (SEM) image of a resulting nanopillar can be found in Fig. \ref{pillar_g2_fig}(a).

We verify that we have isolated single emitters in the diamond nanopillars by performing second-order autocorrelation ($g^{(2)}[\tau]$) measurements. A measurement of $g^{(2)}[\tau]$ for an emitter in a nanopillar under 1-mW, 532-nm excitation is presented in Fig. \ref{pillar_g2_fig}(b). The data are fit to the function 
\begin{equation}\label{g2_eqn}
    g^{(2)}[\tau]=1-c\left((1+b )e^{-\left|\tau\right|/\tau_1}-b e^{-\left|\tau\right|/\tau_2}\right),
\end{equation}
where $b$, $c$, $\tau_1$, and $\tau_2$ are fitting parameters \cite{g2book}. $\tau_1$ provides an estimate of the excited state lifetime of the emitter. $\tau_2$ is the decay time on the photon bunching and relates to a third shelving state in the system. This third shelving state has been discussed for SnV$^{-}$ by Iwasaki \textit{et al.} \cite{IwasakiSnV} and explored theoretically and experimentally for SiV$^-$ \cite{NeuSiVQE,GaliMazePRB2013}.
For the emitter presented in Fig. \ref{pillar_g2_fig}(b), $g^{(2)}[0]=0.23\pm 0.02$, indicating that this emitter is indeed a single emitter. $\tau_1$ and $\tau_2$ were found to be $4.8\pm 0.1$~ns and $103\pm 10.$~ns, respectively.
The estimated emitter lifetime $\tau_1$ of $4.8$~ns is on par with previous measurements of $\sim5$~ns  \cite{IwasakiSnV}. 

Based on a survey of 200 pillars designed for a diameter of 200~nm, we estimate that $2\%$ of such pillars contain low-strain single emitters. This estimate is based on the number of pillars with ground-sate splitting $\sim850$~GHz and $g^{(2)}[0]<0.5$. About $4\%$ of 200-nm pillars contained single SnV$^{-}$, strained or unstrained. Additionally, by counting the number of SnV$^{-}$ in 200-nm diameter pillars and comparing the areal density to the implantation dose, we estimate the conversion efficiency to be $0.7\%$. This conversion efficiency may be improved by varying the implantation dose and annealing time.

For all photoluminescence (PL) studies, we use 532-nm laser light to excite the emitter above resonance. We measure our data with two home-built confocal microscopes with free-space paths into cryostats: a Montana Instruments Cryostation and an attoDRY2100. Further details on our setups can be found in the Supplemental Materials \cite{supplement}.

\section{Optical Properties}\label{polarizationsection}
\begin{figure*}[!htbp]
\includegraphics[width=1\textwidth,]{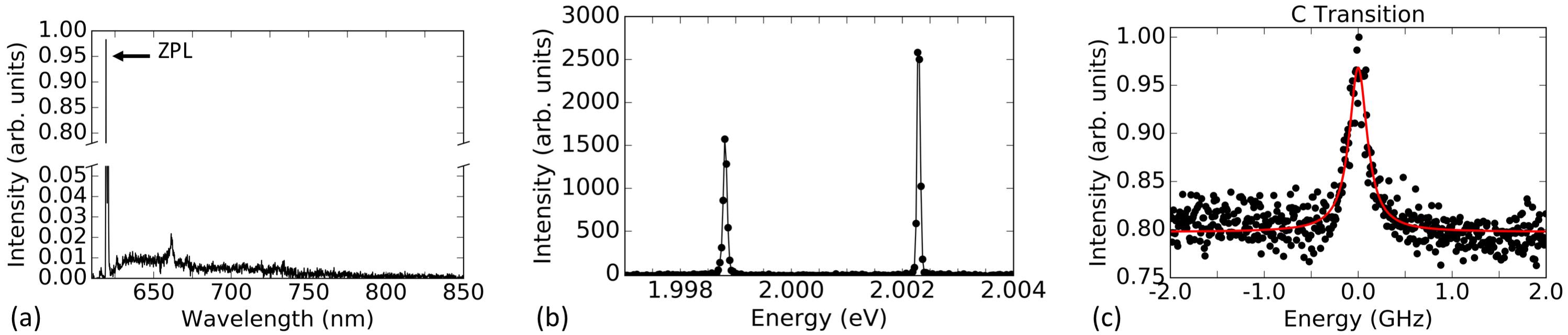} 
\caption{PL and PLE. (a) Broadband PL spectrum of a SnV$^{-}$ taken at 5~K, normalized to the maximum counts on the raw data. (b) PL spectrum of the C and D transitions of a single emitter in a pillar taken at 5~K. Data are presented as black points joined by black lines. (c) PLE scan across the C transition of a SnV$^{-}$, normalized to maximum value of data (black points). A Lorentzian fit to the data is shown in red. The linewidth is found to be $231.9\pm 9.5$~MHz.}
\label{PL_PLE_fig}
\end{figure*}

PL spectra of a single SnV$^{-}$ in a nanopillar at $5$~K are presented in Fig. \ref{PL_PLE_fig}(a) and (b). We observe strong emission into the C and D transitions, which present as two closely spaced, sharp lines around 620~nm in Fig. \ref{PL_PLE_fig}(a). As expected at this temperature, only the C and D transitions are visible \cite{IwasakiSnV}. The phonon sideband is also visible in the background-subtracted plot of Fig. \ref{PL_PLE_fig}(a). The background PL spectrum, which can be found in the Supplemental Materials \cite{supplement}, was acquired with the same excitation power and integration time on another pillar of the same size that did not contain an emitter. Fig. \ref{PL_PLE_fig}(b) is a higher-resolution PL spectrum of the C and D transitions only. From this spectrum the ground-state splitting of the emitter is found to be 842~GHz, close to the $\sim850$~GHz expected, indicating that this emitter is under minimal strain. 

To obtain the linewidth of the emitters, we perform photoluminescence excitation (PLE) experiments. Here, we tune a laser across the ZPL and collect the photons emitted into the phonon sideband. We measured the C transition via PLE, shown in Fig. \ref{PL_PLE_fig}(c), and found the linewidth to be $231.9\pm9.5$~MHz at 1.7~K. The emitter measured for this experiment had a $\tau_1=3.8$~ns based on $g^{(2)}[\tau]$ data, which corresponds to a lifetime-limited linewidth of approximately $42$~MHz. The discrepancy between the lifetime-limited and measured linewidths may be caused by a significant spectral diffusion that we observed during the measurements. The spectral diffusion may be mitigated by implanting with a lower dose.

\begin{figure}[]
\includegraphics[width=0.45\textwidth,]{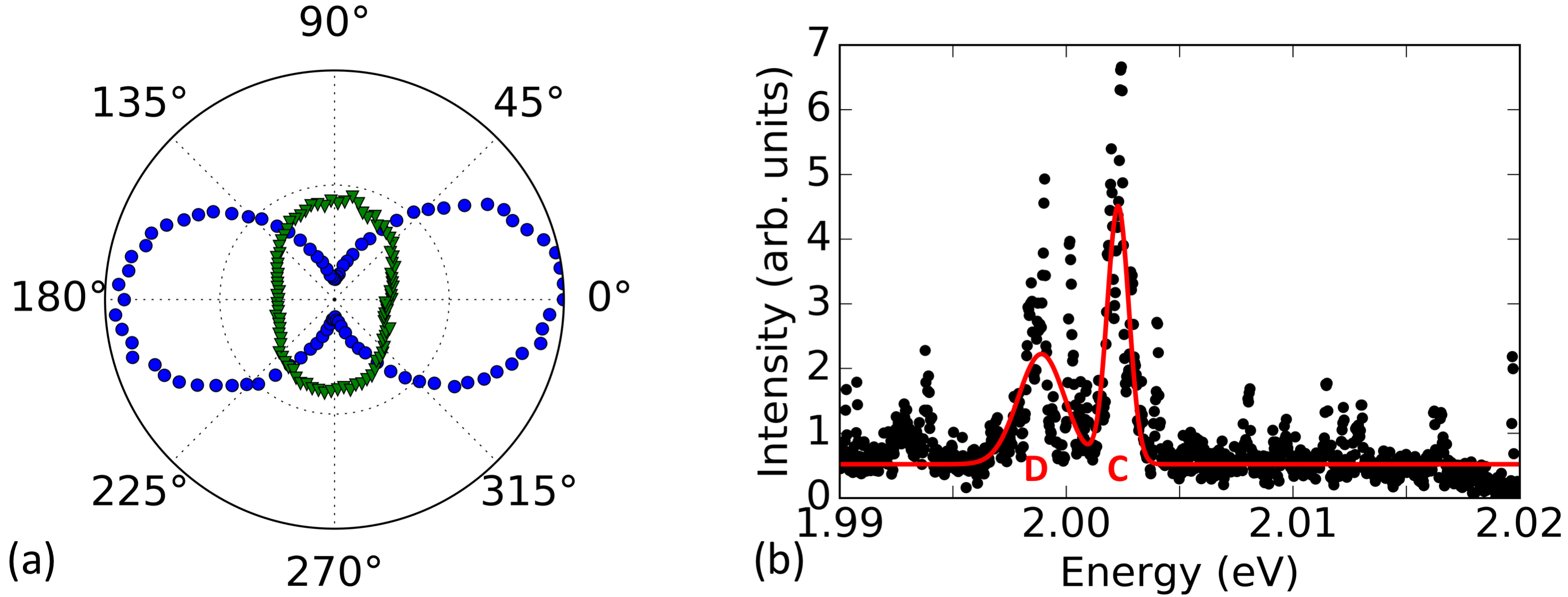} 
\caption{PL spectrum and polarization dependence.  (a)  Polarization dependence of the emission. Intensity of the C (blue dots) and D (green triangles) transitions as a function of angle, normalized to the maximum value of transition C.
(b) Average PL spectrum of 44 pillars. C and D peaks are two distinct, broad peaks on either side of 2.00~eV. The red line is a double Gaussian fit to the two peaks.}
\label{PL_and_pol_fig}
\end{figure}

We also study the polarization of the emission into the C and D transitions for a single SnV$^{-}$. To do so, we use the setup illustrated in Fig. S1 of the Supplemental Materials \cite{supplement} and measure the spectrum with different half-wave plate angles. The C and D transitions are then fit to Lorentzians, the areas under which are taken to be the emission intensity.
The emission intensity of the C and D transitions are plotted as a function of the half-wave plate angle in Fig. \ref{PL_and_pol_fig}(a). The SnV$^{-}$ studied displays a clear polarization dependence in its emission. The strong polarization dependence is, as expected, similar to that of SiV$^{-}$ in diamond \cite{HeppSiVB} and is consistent with previously reported room-temperature absorption data for SnV$^{-}$ \cite{TchernijSnV}. From Fig. \ref{PL_and_pol_fig}(a) we observe that the dipole moments associated with the C and D transitions projected onto the $(001)$ diamond surface are perpendicular to one another.  
 
We estimate the spatial inhomogeneous broadening present in the emitters by plotting the average PL spectra of 44 pillars, as shown in Fig. \ref{PL_and_pol_fig}(b). The C and D transitions appear in the plot as two distinct peaks and have been fit to Gaussian distributions. From the fits, we find the full widths at half-maxima of the C and D transitions to be $271\pm8$~GHz and $583\pm27$~GHz, respectively. Raman tuning has been used to tune the ZPL of a SiV$^{-}$ up to 100 GHz \cite{SipahigilScience2016,SunRamanPRL}. Thus, it may be possible in future works to overcome a significant portion of the inhomogeneous broadening found here. 

\section{Zeeman Splitting}\label{ZeemanSection}
\begin{figure*}[!htbp]
\includegraphics[width=1\textwidth,]{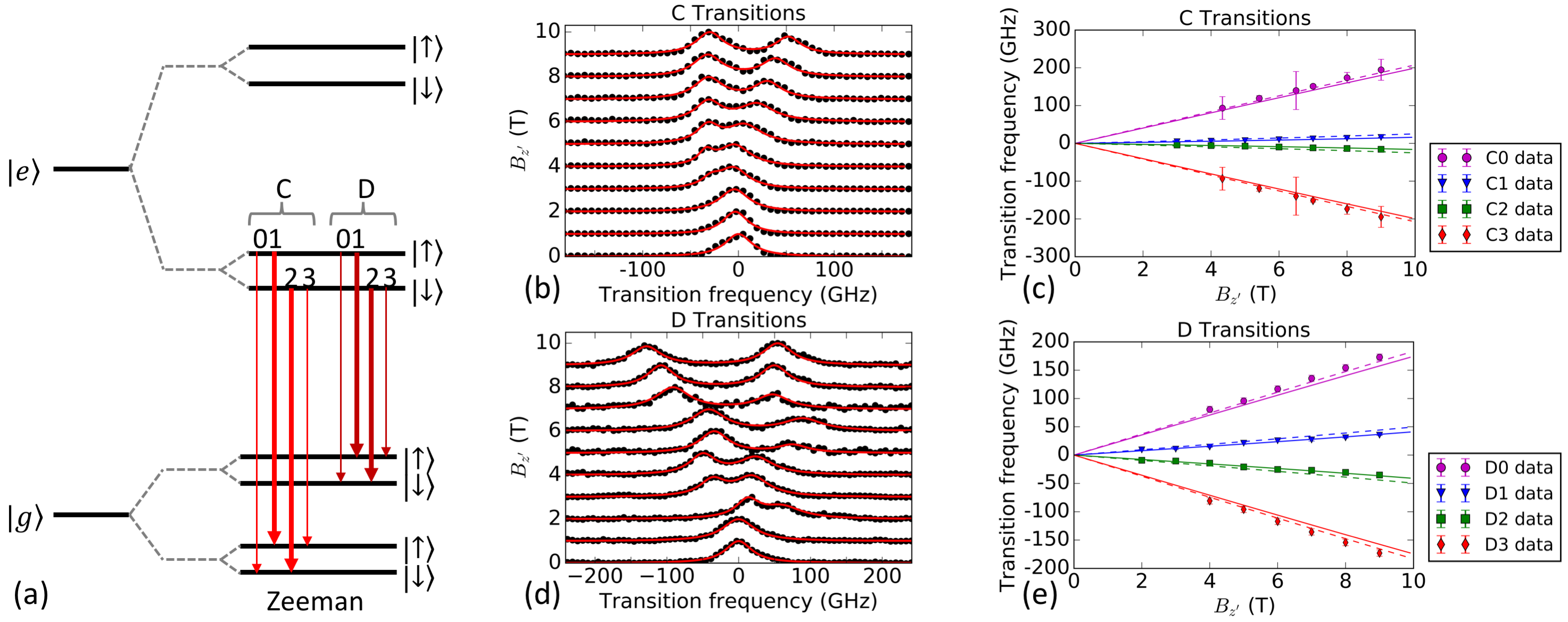} 
\caption{Behavior of SnV$^{-}$ in an external magnetic field $B_{z^{\prime}}$. (a) Schematic energy level diagram of SnV$^{-}$ when a magnetic field is applied. Each of the original optical transitions splits into four transitions. Two of each set of four transitions are spin-conserving and thus strongly allowed (bold arrows), and the other two are weakly allowed (thin arrows). Transitions C (red) and D (dark red) are represented by arrows, numbered 0 to 3 in order of decreasing energy. (b) Stacked spectra of inner transitions C1 and C2 as a function of $B_{z^{\prime}}$ collected with a double monochromator. Transition frequency is with respect to the central transition frequency at 0~T. Solid red lines are double or single Lorentzian fits to the data (black points). (c) Location of central frequencies of all four C0, C1, C2, and C3 transitions. Error bars on the data points are the estimated error based on the variance of central frequency fit parameters estimated by \texttt{curve\_fit} in Python. Solid curves are functions derived theoretically from Eq. \ref{PRX_Ham} and correspond to data of the same color. Dashed curves are functions derived from Eq. \ref{PRX_Ham} when $f^{g,u}$ and $\delta_f^{g,u}$ are scaled by the optimal $\alpha^{g,u}$. (d) Same as (b) for transition D. (e) Same as (c) for D transitions.}
\label{Zeeman_fig}
\end{figure*}

We explore the spin properties of a single SnV$^{-}$ and compare our experimental results to theoretical predictions. When a magnetic field is applied to a SnV$^{-}$ center, the two-fold spin degeneracy is lifted and the two branches of the excited- and ground-state manifolds each split into two more states. Consequently, each of the original optical transitions splits into four: two spin-conserving transitions that are strongly allowed and two spin non-conserving transitions that are only weakly allowed by magnetic field components orthogonal to the symmetry axis of the emitter. A schematic diagram of the energy levels and optical transitions is shown in Fig. \ref{Zeeman_fig}(a).

For our theoretical model, we solve for the eigenstates of the Hamiltonian derived in Ref.~\onlinecite{GaliPRX}:
\begin{widetext}
\begin{equation}\label{PRX_Ham}
    \hat{H}_{eff}^{g,u}=-h\lambda^{g,u}\hat{L}_z\hat{S}_z+\mu_Bf^{g,u}\hat{L}_zB_z+\mu_Bg_S\hat{\textbf{S}}\cdot\textbf{B}+2\mu_B\delta_f^{g,u}S_zB_z+\hat{\Upsilon}_{strain},
\end{equation}
\end{widetext}
where the $z$-direction is along the symmetry axis of the emitter. The superscripts $g$ and $u$ respectively denote the parity of the states: even (\textit{gerade}) and odd (\textit{ungerade}). For SnV$^{-}$ and similar negatively charged group-IV-based color centers, the ground state corresponds to $g$ and excited to $u$. The sign on the effective spin-orbit coupling (first) term opposes that of Hamiltonians previously used to model other negatively charged group-IV-based color centers \cite{GaliPRX,HeppSiVB}. This Hamiltonian can be used to find the energy levels of a SnV$^-$ as a function of applied DC magnetic field \textbf{$B$} for arbitrary magnetic field direction. The third term, which contains $\hat{\textbf{S}}\cdot\textbf{B}$, describes the Zeeman effect.
In the form $x=\{x^g,x^u\}$, where $x$ is a parameter, the values given in Ref. \onlinecite{GaliPRX} for the parameters in Eq. \ref{PRX_Ham} are $\lambda=\{850, 3000\}$ GHz, $f=\{0.154,0.098\}$, and $\delta_f=\{0.014,0.238\}$. We study an emitter under minimal strain, so we neglect the $\hat{\Upsilon}_{strain}$ term.

For the measurements in a magnetic field, our sample is cooled to 1.7 K in an attoDRY2100 that is equipped with a superconducting magnet. The magnetic field is applied out of the plane of the chip, along the $[001]$ direction, which shall be denoted as $z^{\prime}$. We apply magnetic fields $B_{z^{\prime}}$ from 0 T to 9 T, inclusive, and perform PL spectroscopy. To collect spectra of the inner transitions with finer resolution, we use a double monochromator. The resulting double monochromator data as a function of field is presented in Figs. \ref{Zeeman_fig}(b) and (d). A detailed explanation of how the data in Figs. \ref{Zeeman_fig}(c) and (e) were extracted from the raw spectra can be found in the Supplemental Materials \cite{supplement}.

We plot the experimental and theoretical frequencies of the four peaks comprising C and D in Figs. \ref{Zeeman_fig}(c) and (e). To correct for a slow spectral drift, we center the inner and outer frequencies pairwise about their average frequency. We apply the same correction to the theoretical plots. 
As can be seen in Figs. \ref{Zeeman_fig}(c) and (e), the experimental data are in good agreement with the predicted behavior based on Eq. \ref{PRX_Ham}.

We note, however, that two of the values used to generate the solid curves of Figs. \ref{Zeeman_fig}(c) and (e) are dependent on $g_L^{g,u}$ values for SiV$^{-}$.  Specifically, $f^{g,u}=p^{g,u}g_L^{g,u}$ and $\delta_f^{g,u}=\delta_p^{g,u}g_L^{g,u}$, where only the $p^{g,u}$ and $\delta_p^{g,u}$ values were calculated for SnV$^-$ by the authors of Ref. \onlinecite{GaliPRX}, while $g_L^{g,u}$ values were found for SiV$^{-}$. Thus, we have tried scaling the $g_L^{g,u}$ values by a constant $\alpha^{g,u}$ to find the least squares fit of the theory to the data. We have found the best-fit values of $\alpha^{g,u}$ to be $\alpha=\{0.98,1.32\}$. The theoretical curves that include this scaling are shown as dashed curves in Fig. \ref{Zeeman_fig} (c) and (e) and yield an improved agreement between the theory and the data---the sum of the squared residuals is reduced by a factor of 2 compared to the original case. Thus, while the original model is in good agreement with our data, it can be further improved by changing the $g_L^{g,u}$ values used. 

\section{Summary and Outlook}
We have studied the optical properties of single SnV$^{-}$ centers that have been isolated in diamond nanopillars. We were able to generate emitters with narrow linewidths of $\sim232$~MHz. These linewidths are an improvement over previously measured linewidths in bulk diamond and demonstrate that emitters can be incorporated into nanostructures without degrading the optical properties. SnV$^{-}$ linewidths may be further improved with lower implantation dose and a longer high-temperature anneal. 
We also observed a strong polarization dependence of the emission, with the C and D emission being orthogonally polarized. Lastly, we observed the behavior of a single SnV$^{-}$ as a function of magnetic field and found it to be in good agreement with theoretical predictions presented by Thiering and Gali \cite{GaliPRX}. 
Our findings indicate that the SnV$^{-}$ is a promising candidate for a spin qubit and warrants further investigation. Future works include characterizing the spin coherence time $T_2$ \cite{PingaultSiVT2CPTPRL2014,RogersSiVT2CPTPRL2014,PingaultSiVT2ODMRPRL2017} of the isolated emitters and fabricating more sophisticated nanostructures \cite{DoryOptimizedDiamondPhotonics} such as photonic cavities resonant with SnV$^{-}$ ZPLs, two necessary steps in order to implement the SnV$^{-}$ as an optically interfaced spin qubit.

\section{Acknowledgments}
We acknowledge Gerg\H{o} Thiering and \'{A}d\'{a}m Gali for fruitful discussions about theory. We also acknowledge Linda Zhang and Greg Pitner for helpful advice on fabrication. Funding: This work is financially supported by Army Research Office (ARO) (award no. W911NF1310309), National Science Foundation (NSF) RAISE TAQS (award no. 1838976n), and Air Force Office of Scientific Research (AFOSR) DURIP. A.E.R. acknowledges support from the National Defense Science and Engineering Graduate (NDSEG) Fellowship Program, sponsored by the Air Force Research Laboratory (AFRL), the Office of Naval Research (ONR) and the Army Research Office (ARO). C.D. acknowledges support from the Andreas Bechtolsheim Stanford Graduate Fellowship. Part of this work was performed at the Stanford Nanofabrication Facility (SNF) and the Stanford Nano Shared Facilities (SNSF), supported by the National Science Foundation under award ECCS-1542152.

At the time of submission of this manuscript, we became aware of a very recent, similar work \cite{EnglundSnV}.


%

\end{document}